# Neutral multi-MeV/u particles from laser-induced processes in ultra-dense deuterium D(0): accurate two-collector timing and magnetic analysis

Leif Holmlid
Department of Chemistry and Molecular Biology, University of Gothenburg, SE-412 96 Göteborg, Sweden
Email holmlid@chem.gu.se

**Abstract**

Laser-induced processes in ultra-dense deuterium D(0) layers eject multi-MeV $u^{-1}$ particles using ns laser pulse energies of <200 mJ. Such particles have been observed previously as mA currents to time-of-flight (TOF) collectors at up to 1 m distance. The signal current is mainly due to the ejection of secondary electrons by impinging MeV particles on the collectors. Improved two-collector time-of-flight measurements now show that the energy of the particles is in the range 1-50 MeV $u^{-1}$. Their distributions are almost thermal at up to 13 MeV $u^{-1}$ or are sharper than thermal. The fastest sharp peak may indicate shock-wave acceleration by many-body energy transfer. A magnetic field of 0.4 T deflects only a small part of the multi-MeV particle flux which thus mainly consists of neutral particles. By combining the TOF method with magnetic deflection, it is ascertained that the multi-MeV particles are studied and not any slower particle emission from the target. The neutral multi-MeV particles are concluded to be fragments of ultra-dense hydrogen $H_N(0)$ as observed in other experiments.

## 1. Introduction

Two ultra-dense hydrogen materials have been proved to exist so far, ultra-dense deuterium D(0) [1,2] and ultra-dense protium p(0) [2,3]. They both exist in a few different forms with slightly different bond distances and densities. All these forms may be characterized as spin-based Rydberg Matter (RM) [2]. This description is based on a theory developed by J.E. Hirsch [4]. The typical inter-nuclear distance is 2.3 ± 0.1 pm as found for D(0) in $s=2$ in several studies [1,5,6]. This corresponds to a density of $10^{29}$ cm$^{-3}$ or >140 kg cm$^{-3}$. This density is a factor of $10^5$ higher than in ordinary condensed hydrogen. It is obvious [7,8] that such a density will make it relatively easy to produce nuclear fusion in these materials, at least in D(0). There, normal D+D fusion processes giving $^4$He and $^3$He ions is shown to be initiated by a relatively weak pulsed laser [9].

The ejection of particles with many MeV u$^{-1}$ energy from laser-induced processes in D(0) as well as from p(0) has been reported in several publications [10-18]. Such particles indicate nuclear fusion, either as a source or as a result. These particles have been studied by time-of-flight (TOF) to one or several detectors or collectors in-line, using delays in metal foils [14] and simultaneous TOF to two collectors [15,16] to ascertain that massive particles are observed. The question which has aroused most interest is the charge state of these massive particles: for charged particles, the standard method of analysis is a Thomson parabola study to identify the ions and their typical energies. Since for D(0) most of the MeV particles are neutral, the Thomson parabola method [19,20] is not adequate and the present study employs a combination of magnetic deflection and TOF methods to make sure that the MeV particles are studied and not any slower particles. Further, recent studies show that the neutral particles ejected decay within a few ns to charged or neutral particles with even higher kinetic energy [18]. The main problem that is addressed here is the charge and nature of the particles ejected by the laser-induced processes in the ultra-dense materials. It is apparent that the detection of large fluxes of particles with energies up to 20 MeV u$^{-1}$ indicates nuclear fusion processes, either as the <u>source</u> of these particles or as a <u>result</u> of these particles.

## 2. Theoretical background

Ultra-dense deuterium D(0) is the lowest energy form of deuterium atoms. This material was previously named D(-1) since it was thought to be an inverted form of matter where nuclei and electrons had exchanged roles. Since it now is clear that the description as a spin-based form of Rydberg matter [2] agrees well with experimental results, the angular momentum is now used as the parameter as for other types of Rydberg matter [21]. Several forms exist, but the most common form has a D-D bond distance of 2.3 ± 0.1 pm [1,5,6]. D(0) is closely related to dense deuterium D(1) which has a D-D distance of 150 pm. More background is given in a recent review on Rydberg matter [21]. The microscopic structure of D(0) is given by chain clusters $D_{2N}$ with *N* integer. Such clusters contain pairs of nuclei which rotate around the axis in the cluster [22]. The ultra-dense materials are quantum materials with properties like superfluidity and superconductivity [23]. These properties were verified for D(0) by a fountain effect [24] and a Meissner effect [25], both observed at room temperature. The superfluid nature of both D(0) and p(0) is directly observed in experiments which use this property to form thin liquid films on metal surfaces [3,26]. It is found that condensation to D(0) does not take place on organic and inorganic polymer surfaces [14,27]. The similarity of D(0) to other superfluids was first discussed by Winterberg [7,8].

The theory for the experimental proof of the extremely short bond distances in D(0) is summarized here. When a chemical bond is broken by photons for example by a relatively weak laser beam as used here, the electrons in the bond are either excited to higher energies (orbitals) or are ejected from the bond, leaving one or several ions behind. The excess energy given to the ionic fragments depends on their distance when they are formed by the electron removal. If this excess energy is larger than the bond energy to other atoms in the material, the ion (atomic or molecular) may be ejected from the material. This process is a form of Coulomb explosion (CE). The maximum energy release is easily calculated from the initial distance between the charges when they are formed (thus when the electrons are removed). Thus, it is possible to determine the initial repulsion energy between the ions by measuring the kinetic energy of the fragments at a large enough distance. The distance between the ions before the break-up is found directly from the Coulomb formula as

$$r = \frac{1}{4\pi\varepsilon_0} \frac{e^2}{E_{kin}} \qquad (1)$$

where $\varepsilon_0$ is the vacuum permittivity, $e$ the unit charge and $E_{kin}$ the sum kinetic energy for the two fragments (KER) from the CE. The fraction of the KER that is observed on each fragment depends of course on the mass ratio of the fragments. The kinetic energy is determined most easily by measuring the TOF of the particles and converting this quantity to kinetic energy. This requires that the mass of the particle is known or can be inferred, which is relatively easy in the case of deuterium. If the distance between the ions is of the order of 100 pm as in short chemical bonds, the excess energy released is 14 eV to be shared between the two ions, thus maximum 14 eV on the light fragment in the explosion (minus bond energies). Now, the normal kinetic energy (velocity) from laser-induced CE determined by time-of-flight in numerous experiments is 315 eV u$^{-1}$ [1,5,6]. The fastest particles observed when working with deuterium have reasonably a mass of 2, thus a kinetic energy of 630 eV. This means an initial distance of 2.3 pm between the ions. This distance is a few percent of a normal chemical bond, which means that $(100/2.3)^3 = 8\times10^4$ gives the order of the density of this ultra-dense material relative to ordinary materials, or a density of the order of 100 tons per dm$^3$. It should be observed that the energy of the fragments generated by the CE does not change with the laser intensity. This means that any acceleration of the ions due to the laser is excluded, which is also known from the low laser intensity used.

The laser ejected particles from D(0) are neutral as shown here and they probably consist of cluster fragments of D(0) and p(0), both described as H$_N$(0). The ejection is apparently in the form of a sheath which is accelerated by the laser-initiated release of energy. In published experiments [14,15], a large energy gain is observed, with a higher total energy given to the MeV particles than introduced by the laser pulse. This observation points to nuclear fusion as the source of the excess energy. The importance of H(0) also in other laser-induced fusion experiments was recently described [28].

## 3. Experimental

A Nd:YAG laser with pulse energy of < 0.2 J was used, with 5 ns pulses at 532 nm and normally 10 Hz repetition rate. The source used for producing D(0) is described in the literature [5]. In the source, a potassium doped iron oxide catalyst sample [29,30] forms D(0) from deuterium gas (99.8% $D_2$). The D(0) formed falls down to the target plate below the source where the laser beam impinges. The direction of the flux to the collectors is at approximately 60° towards the normal of the target and at 45° relative to the laser beam.

The set-up used is shown schematically in Fig. 1 [1,12]. The vacuum chamber has a base pressure $< 1 \times 10^{-6}$ mbar. The $D_2$ gas pressure in the chamber is up to $1 \times 10^{-5}$ mbar (uncorrected ionization gauge reading). On the target plate which is sloping at 45° angle towards the horizontal, a foil of Ta is mounted. The laser beam is focused on the foil with an $f = 400$ mm spherical lens. The intensity in the beam waist of (nominally) 30 µm diameter is relatively low, $\leq 3 \times 10^{12}$ W cm$^{-2}$ as calculated for a Gaussian beam.

In the wall of the cylindrical chamber in Fig. 1 at a distance of 22 cm from the target, an aperture with 5 mm diameter limits the angular spread of the particle flux. In the two-collector experiments, a foil collector at 64 cm distance from the target is used as the inner collector. It can be rotated out from the beam to the outer collector. This inner collector has a diameter of 53 mm while an annular grounded plate in front of it has an opening diameter of 38 mm. The outer foil collector is at 103 cm distance from the target. The signal is measured directly with a 300 MHz digital oscilloscope with a 50 Ω coaxial input. Battery bias of -24 V can be applied to the collectors still with 50 Ω coaxial input. A diode at the laser gives the trigger signal. The length of the cabling is adjusted so that the maximum zero point error in the trigger signal is 3 ns at the oscilloscope.

In the magnetic deflection experiments, the flux ejected from the target is analyzed by a magnetic field provided by four small permanent cylindrical magnets (neodymium magnets with diameter 14 mm, height 8 mm) in pairs above and below the beam to the collectors. The magnets were mounted on an arm which could bring them in and out from the beam to the collectors. The distance between the pole faces in each pair is 5 mm. The magnetic field strength has been measured with a Hall effect sensor (Allegro A1326 giving 2.5 mV G$^{-1}$). The field strength at a distance of 8 mm from the pole of one magnet was 0.1 T, and the field between the magnets in the beam was > 0.4 T. The total length at this field strength along the beam was approximately 28 mm. Just behind the magnets, a central opening of 4×4 mm on the arm was used to define the beam. Further in the beam direction at 2 cm distance, a defining slit with width 0.8 mm and vertical height 5 mm was mounted in the wall between the central chamber and the separately pumped collector chamber, at 22 cm distance from the target. In the magnet experiments, the inner collector was in the form of a vertical pin of 50 mm length and 1 mm diameter which could be moved sideways by rotating the flange on which it was mounted with an offset as shown in Fig. 1. The distance from the target was 64 cm. A laser pointer on the same flange was used to observe the rotation of the flange and thus the sideways motion of the pin. No large signal was observed outside ± 3 mm sideways motion relative to the central beam. The pin current is measured directly with a 300 MHz digital oscilloscope with a 50 Ω coaxial input, using -24 V battery bias on the pin. Thus, no electrons due to photons from the target can reach the pin. The outer collector was used to optimize the adjustment of the laser on the target in the magnetic deflection experiments.

Neutron emission in the high-pressure chamber is monitored by using bubble detectors (Bubble Technology Industries, BTI) types BN-PND for high energy (<15 MeV) and BDT for thermal neutrons. They are located outside the vacuum wall of the chamber. Typically 1-2 bubbles are formed during each experimental run, corresponding to approximately 0.2 - 0.5 µSv h$^{-1}$.

## 4. Results

### 4.1. Multi-MeV particles

Two-collector experiments have been performed in numerous runs. They show that the normal result is a typical delay between the collectors, corresponding to the particle distance travelled, thus proving that massive non-relativistic particles are observed (and not photons). The signal observed varies with the applied bias to the collector, as further described in [15,16].

The TOF signal observed at zero collector bias in all figures shown here is due to particles with energy 1-20 MeV u$^{-1}$. This has also been demonstrated previously by TOF collector measurements [14] and two-collector TOF measurements [15,16], but the present measurements are more exact with sharper TOF peaks. The signal observed at each collector is due to the secondary electrons ejected from the collector by the impinging massive particles. A contribution may also exist from pair production [17]. With zero bias, the signal observed is positive since electrons are ejected from the collector by the impinging particles. The secondary emission coefficient for MeV protons and deuterons is known [31-33]. A negative bias on the collector decreases the space charge at the collector and thus extracts a larger fraction of the low-energy secondaries, giving a higher positive signal to the collector but also worse time resolution for this reason [15,16]. Also a signal due to photons from the plasma may be observed in experiments with negative bias [15,16]. Thus, mainly results with zero bias are discussed here due to the better resolution. A positive bias is more difficult to use beneficially, since it tends to attract secondary electrons from the surrounding apparatus walls, giving a slow negative signal at the collector. Also, thermal electrons from the plasma are observed at long times (ms) with positive bias, but they do not reach the collector when it is at zero bias [15]. Thus, the results given here are mainly measured with zero bias, possibly giving somewhat fewer electrons than the number of impinging particles (nuclei) and thus a slightly lower signal than corresponding to the number of particles (nuclei). An absolute normalization is however not required here.

The results in Figs. 2 and 3 show directly the delay of the signal caused by the distance between the collectors. In the upper panel in the figures, the directly measured signal is shown, while the lower panels show the time-base for the outer collector recalculated by the ratio of the distances 64/103, thus bringing the signals in agreement at the distance of the inner collector. It is possible to calculate the particle energy directly from the transit time between the collectors, and this gives an improved value for the zero time in the TOF spectra with a 3 ns shift of both curves. Note that the inner collector was rotated out of the beam for the measurement at the outer collector. Fig. 3 shows the result with an inner slit of 3 mm width at a distance of 74 mm from the laser target (inner slit box in Fig. 1), giving a slightly lower intensity but the same timing performance. Thus, the signal observed is not due to any scattering or channeling effects. An area of a few mm width on the laser target is however seen by the collectors even with this inner slit in place. These results are of interest also since they show two different features, an early peak centered at 50 MeV $u^{-1}$ and a later bump at 1-30 MeV $u^{-1}$. In Fig. 2, thermal distributions are included for all curves using the correct velocities derived from the collector distances. Note that in the top panel, the temperature values are slightly different for the two curves, due to the difference in zero timing which is corrected by the 3 ns shift in the lower panel. In Figs. 2 and 3, the upper panels show directly that even the first peak is broader at the outer collector than at the inner collector, thus proving that this signal peak is due to particles since a photon peak would not show such a velocity distribution. This first peak is sharper (falling faster) than a thermal distribution with the same peak position. The first peak may be interpreted as the direct peak from the explosive ejection from the target, while the second peak is thermal due to scattering events in the emitting plasma. This is discussed further below.

### 4.2. Neutral particles

To study the charge state of the particles observed with energy 1-20 MeV $u^{-1}$, magnetic deflection experiments have been done. The deflection of the particle flux was measured with the movable pin collector shown in Fig. 1. To certify that the same MeV particle signal was measured as in the experiments described above, the signals to the outer collector and the pin collector are compared in Fig. 4, using negative bias. Due to the small slit of 0.8 mm width, a slightly different view of the target region is provided in this experiment than in the other two-collector experiments in Figs. 2-3. Recalculating the time-base for the outer collector gives good agreement between the two curves in the lower panel, which shows that massive MeV particles are observed. Both signal distributions agree well with the 9 MeV $u^{-1}$ theoretical thermal distribution in Fig. 4. The oscillations in the signal to the outer collector are reproducible and may be due to bad impedance matching in the internal conductor structure.

The main result shown in Figs. 5-6 is that most MeV particles are not deflected by the magnetic field outside the angular range studied. This shows that they are not atomic ions p or D$^+$ (see next paragraph for calculated deflections). This can be observed by comparing the signal for D(0) in Fig. 5 without magnet in place with the signal in Fig. 6 with the magnetic field. The difference between these two sets of data is shown in Fig. 7. The sum of the displayed signal values in Fig. 5 is 35.3 (in a volts scale) while that in Fig. 6 is 26.3. This means that 25% of the signal is lost with the magnet in the beam. This is probably an effect of the limiting slits, and could be compensated for in the difference displayed in Fig. 7, by increasing the signal with the magnet in the beam a corresponding amount. No such flux is detected after deflection in the magnetic field. The signal difference shown in Fig. 7 indicates a deflection of a small part of the signal from D(0) from negative to positive deflections, as expected for a positive ion flux. The deflection is of the order of 2 mm and the total flux deflected in this way is estimated to be 10% of the total flux without magnet.

Since the velocity distributions of the particles are broad from 1 to at least 20 MeV u$^{-1}$, a well-defined deflection pattern is not expected. A calculation of the deflection of a proton with 12 MeV gives 11 mm, and for 1 MeV 37 mm, thus far outside the deflection range studied in Figs. 5-7. The deflections observed are 2 mm as concluded above, with no observable difference between slow and fast particles. With a mass-to-charge ratio of 20, the deflection at 12 MeV u$^{-1}$ energy is only 0.5 mm, and at 1 MeV u$^{-1}$ it is 1.6 mm. This ion result could possibly indicate low energy carbon ions from a hydrocarbon layer on the target. This is however unlikely due to the continuous renewal of the superfluid D(0) layer on the target from the D(0) source. The results indicate that the particles observed to deflect are clusters p$_{20}$(0)$^+$ or D$_{10}$(0)$^+$, or mixed forms H$_N$(0)$^+$. It is concluded that the magnetic deflection experiments show that 90% of the particles detected are neutral clusters H$_N$(0).

## 5. Discussion

### 5.1. Fast peak at 50 MeV u$^{-1}$

In Figs. 2 and 3 a fast TOF peak is observed. This peak is often observed from D(0) on various carrier materials like Ta [14] but often not so well separated from the remaining distribution as in the present case. This means that further information can be extracted here. This peak has previously been suggested to be due to photons and possibly to fast electrons. It is apparent that it is not due to photons since the peaks at the two collectors do not agree after a 1.3 ns shift to compensate for the velocity of light. This is also apparent from the broadening of the peak at the outer collector, which indicates a velocity distribution of massive particles. Calculating the velocity of this peak by comparing the two collector signals gives approximately 30% of the velocity of light, or 50 MeV u$^{-1}$ (without relativity correction). Thus it is due to fast massive particles. Fast electrons (or positrons) with 25 keV energy would have a 6.5 ns TOF matching the peak observed in Figs. 2-3. Such particles would be deflected by 5 cm in the geomagnetic field on the way to the outer collector, thus probably reaching the wall in the TOF tube. No signal change was observed from placing strong permanent magnets at a distance of 2 cm from the beam. Thus, the particles observed are not electrons or positrons. It is observed that this peak is not clearly visible when a limiting slit is used in the beam, like in Fig. 4. This indicates that these particles in the fast peak are ejected from the edges of the plasma region on the target. Assuming that a sheath of material is ejected by the laser pulse, the thinner edges of the sheath may move linearly without collisions, especially when observed at 60º from the normal of the target. The particles from the thicker central part of the sheath (in the plasma blob) may instead form a broader thermal distribution with typical energy of 5-13 MeV u$^{-1}$. The good agreement with the thermal distributions included in Figs. 2-4 supports this view that the broad distributions are formed by internal collisions in the shock-wave region leaving the target, with a temperature in the MeV range. If the broadening of the distributions instead was due to collisions with the residual gas, particles slower than the thermal distributions would be produced giving more tailing than thermal.

An important question is then: what kind of energy release can give a sheath of material with this large number of particles at an average velocity corresponding to 8 MeV u$^{-1}$, which is the kinetic energy of the peak in the TOF distributions? Since the total energy release to particles is large, nuclear processes are most likely. Assuming the total normal fusion process 3 D → $^4$He + p + n as observed from laser-induced fusion in D(0) [9], this means a total energy given off of 21.6 MeV. However, the $^4$He nuclei will probably not interact strongly with the H(0) phase which is blown off and they will probably carry only the 3.5-3.6 MeV given to them by the fusion process [9]. Thus, the two light masses p and n may receive a total energy of approximately 14+3 MeV at average energy 8.5 MeV u$^{-1}$. Thus, the observed peaks and distributions close to 8 MeV u$^{-1}$ in Figs. 2-4 agree well with the expected average particle energy in such a sheath acceleration process. If the total energy of 21.6 MeV instead is released in a complex of six nucleons (3 D) with one of them carrying away the excess energy from the complex, its energy will be 18 MeV. The fast peak at 50 MeV u$^{-1}$ is however not easy to explain in such a separate-particle transport model but requires a many-particle process in the form of a shock wave from an explosion in the ultra-dense D(0) layer on the target.

## 5.2. Magnetic deflection

The magnetic deflection experiments in Figs. 5-7 show clearly that most of the flux is neutral and that only a fraction of the particles is in positive ion form. Experiments with weaker magnetic fields have been done, and no deflected signal due to small atomic ions has been observed. Of course, a small signal of this type with a large energy spread may be difficult to detect here, so it cannot be excluded that such a signal can be observed with other equipment. The difference distribution in Fig. 7 has been repeated in later experiments. It is also interesting to determine the emitting area on the target. The half-width measured with the pin collector at 64 cm from the target is approximately 4 mm. This means that the half-width of the emitting plasma on the target is slightly less than 2 mm. The visible plasma is considerably larger, but it has previously been observed [12,15] that the visible D(0) plasma itself mainly emits light and gamma radiation, while the particles are ejected from the laser focus.

In many cases it is assumed and demonstrated [34,35] that ions observed in high-intensity laser-induced plasma experiments are ordinary small ions like carbon ions, with the source of carbon being hydrocarbons on contaminated surfaces in the vacuum chamber. The results here do not agree with such artifacts. For example, no well defined ions with fixed mass-to-charge ratio are observed and the observed particle energy up to 20 MeV u$^{-1}$ is much too large for high-energy laser production processes at the present low laser intensity [35]. Of course, the sheath acceleration process advocated here might suggest that carbon ions embedded in the H(0) or D(0) sheath will be accelerated to the same velocity as the sheath clusters. The continuous renewal of the D(0) layer on the target from the source makes it unlikely that any hydrocarbons at all exist in the layer. Further, the base pressure in this chamber is low. It is likly that the relatively few ions observed to deflect with mass-to-charge ratio of the order of 20 are hydrogen cluster ions H$_N$(0)$^+$.

### 5.3. Neutral particles

The timing of the collector signals in Figs. 2 – 4 shows clearly that massive particles with energies 1-20 MeV u$^{-1}$ are ejected from the laser target. The magnetic deflection experiments show that most of the particles are neutral, probably being clusters of the form H$_N$(0) [3]. Neutrons are excluded as the particles detected, as can be concluded from the weak penetration properties of the signal-generating particles. Since the multi-MeV particles indicate fusion or other high-energy nuclear processes, it is also necessary to discuss the possibility that particles like mesons are formed in the experiments. A TOF transit to the collectors takes > 10 ns for the signals observed. If a short-lived particle decays on the way to the collector, the fragments will move out of the direct beam and will not be detected. Thus, only long-lived particles can give the signals observed. Also the magnetic experiments are conclusive, since the deflection takes place at > 5 ns after the ejection from the target plasma. Thus, the most of the particles must still be neutral after this time. Not so many types of neutral nuclear particles are known that will still interact with the Coulomb force to give the signal measured at impact on the collectors. The behavior observed in the experiments indicates neutral particles containing both positive and negative fundamental particles, thus neutral clusters of H(0).

It is also necessary to discuss the formation of neutrons in the laser-induced processes. We have used several different standard methods for neutron detection, and only small fluxes of neutrons have been observed. This is probably due to the high density of D(0), as described in other publications [9,12]. However, it is expected that the neutrons and protons have similar initial TOF distributions, since they are formed in a similar way in the fusion process. Protons may be slowed down faster than the neutrons by collisions during the plasma expansion. Since protons are found to have energies up to 50 MeV u$^{-1}$, it is expected that any neutrons will have similar or higher kinetic energy. This means that ordinary methods for neutron detection like bubble meters (useable for neutrons up to 15 MeV) will not be able to detect such neutrons with high probability. Detectors using $^3$He and BF$_3$ require that fast neutrons are slowed down to low energy by collisions in materials like a paraffin or plastic, so that the low-energy neutrons can be observed. Thus, these instruments are normally only specified up to 17-20 MeV and may not work reliably for 50 MeV neutrons.

## 5.4. Many-body processes

The main result here is that most particles with multi-MeV energies are neutral. Also, the few charged particles ejected have broad velocity and mass-to-charge ratio distributions. This result is not unexpected due to the properties of the ultra-dense hydrogen H(0), with strong short bonds in the cluster structure. The results indicate that a sheath of the material is ejected with intact clusters, and that a separate-particle picture for the ejection is not valid. Instead, the laser-induced energy-releasing processes transfer energy to the ultra-dense clusters in many-body collisions. Collisions between separate nuclei play a minor role. This means that characteristic energies from well-defined processes like ordinary nuclear fusion processes will not always be observed, since the energy release is into many degrees of freedom in the ultra-dense material. These results indicate that a moving sheath (expanding blob) with a velocity close to $9\times10^7$ m s$^{-1}$ (50 MeV u$^{-1}$) ejects the particles with almost the same velocity, independent of their mass. This implies that the layer of D(0) which is fusing is several atoms thick, thus it is a layer of D$_N$(0) clusters [5,6,26].

## 6. Conclusions

Fast particles from laser-induced processes in D(0) are now analyzed by time-of-flight to two collectors in-line and by deflection in a strong magnetic field of 0.4 T. The particles have energy in the range 1-50 MeV u$^{-1}$. Most of them exist in thermal distributions with typical energies of 5-13 MeV u$^{-1}$. Most of the particles are neutral, probably in the form of small neutral ultra-dense hydrogen clusters H$_N$(0). Neutrons are excluded. Appoximately 10% of the total flux is deflected in the magnetic field as positive particles with a mass-to-charge ratio around 20. The large energy in the fast particles indicates many-particle collisions for the ejection of a shock-wave or sheath from the ultra-dense cluster layer.

# Figure captions

Fig. 1. Horizontal cut through the apparatus, with various parts indicated. The inner pin collector is at 64 cm distance from the target. It is a 1 mm pin which is mounted on a rotatable flange. The offset on this flange is 26 mm so deflections ± 20 mm are detectable. In some experiments, the pin collector is replaced with a plate (foil) collector. The outer collector is at 103 cm from the target.

Fig. 2. Simultaneous time-of-flight signals to inner and outer collectors. Upper panel shows the direct raw oscilloscope signals, lower panel with time scale for outer collector recalculated by 64/103 (distance ratio). Zero bias, 3 ns shift from the overall matching in the lower panel. Thermal distributions for protons are included in both panels

Fig. 3. Simultaneous time-of-flight signals to inner and outer collectors in the low-pressure chamber with the inner slit box in the beam. Upper panel shows the direct raw oscilloscope signals, lower panel with time scale for outer collector recalculated by 64/103 (distance ratio). Zero bias, 3 ns shift from the overall matching in the lower panel. A thermal distribution at 12 MeV for protons is included in the lower panel.

Fig. 4. Simultaneous time-of-flight signals to pin collector (maximum position) and outer collector with -24 V bias in the low-pressure chamber. Upper panel shows the raw direct oscilloscope signals, lower panel with time scale for outer collector recalculated by 64/103 (distance ratio). A thermal distribution at 9 MeV for protons is included in the lower panel.

Fig. 5. The time-of-flight signal to the pin collector from laser-induced processes in D(0), as a function of sideways position of the pin. No magnet in the beam. Sum of signal values 35.3 (V).

Fig. 6. The time-of-flight signal to the pin collector from laser-induced processes in D(0), as a function of sideways position of the pin. Magnet with field strength 0.4 T in the beam. Sum of signal values 26.3 (V).

Fig. 7. Difference between the D(0) TOF signals in Figs. 5 and 6. The vertical (signal) scale is the same as in Fig. 6 for convenience. The valley to the left is the range where the signal is higher with magnet, and the arrow at the back indicates how the particles have been deflected by the magnetic field.

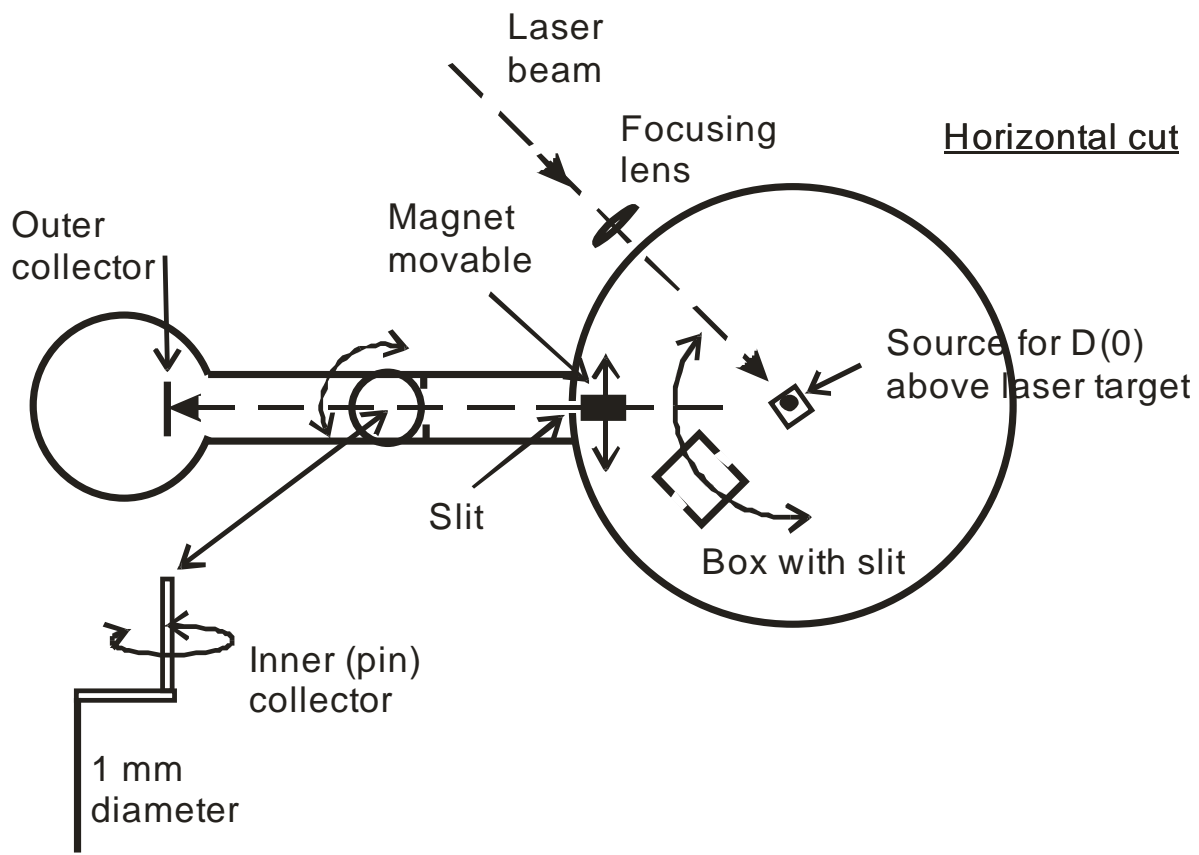

Fig. 1

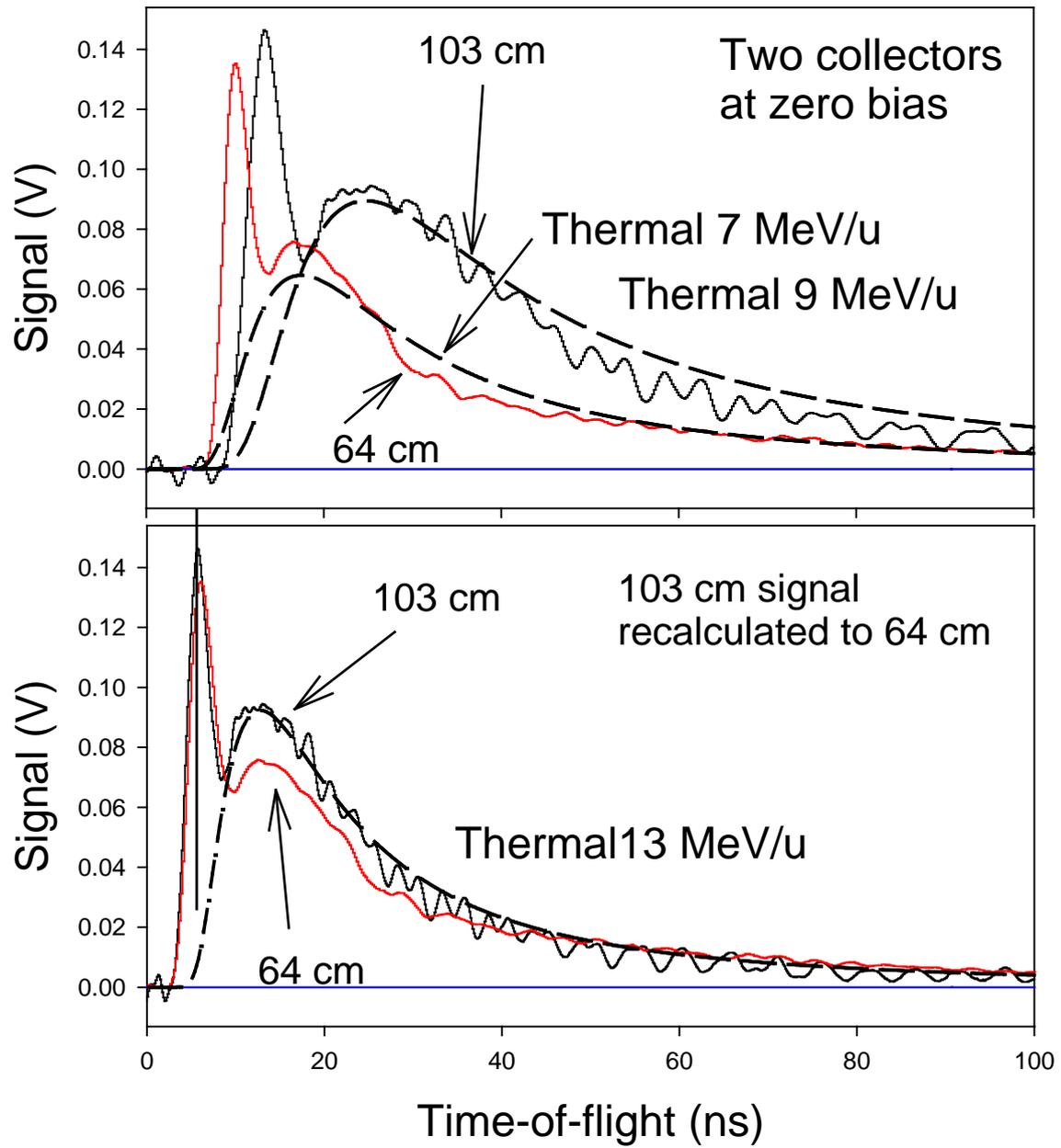

Fig. 2

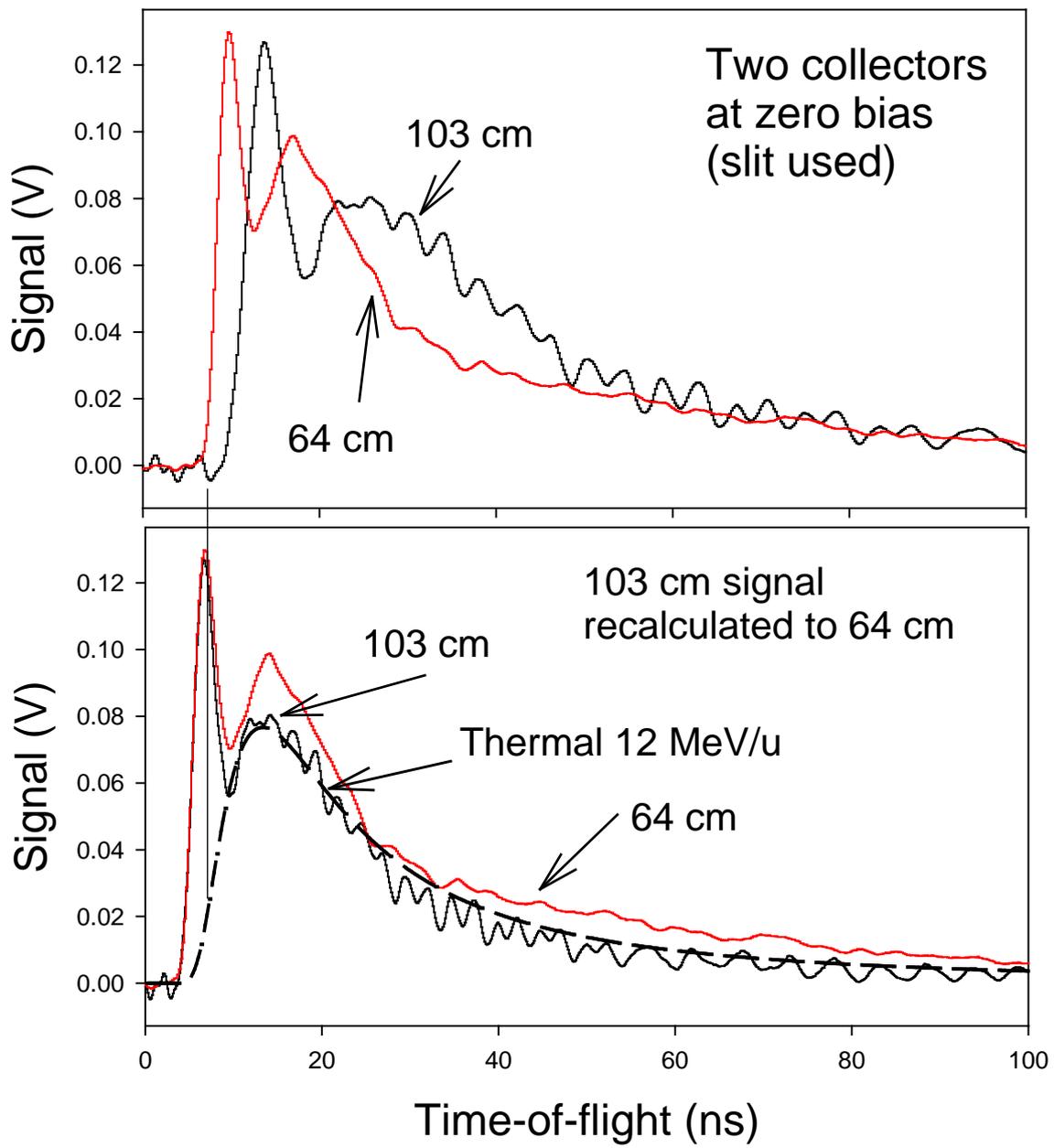

Fig. 3

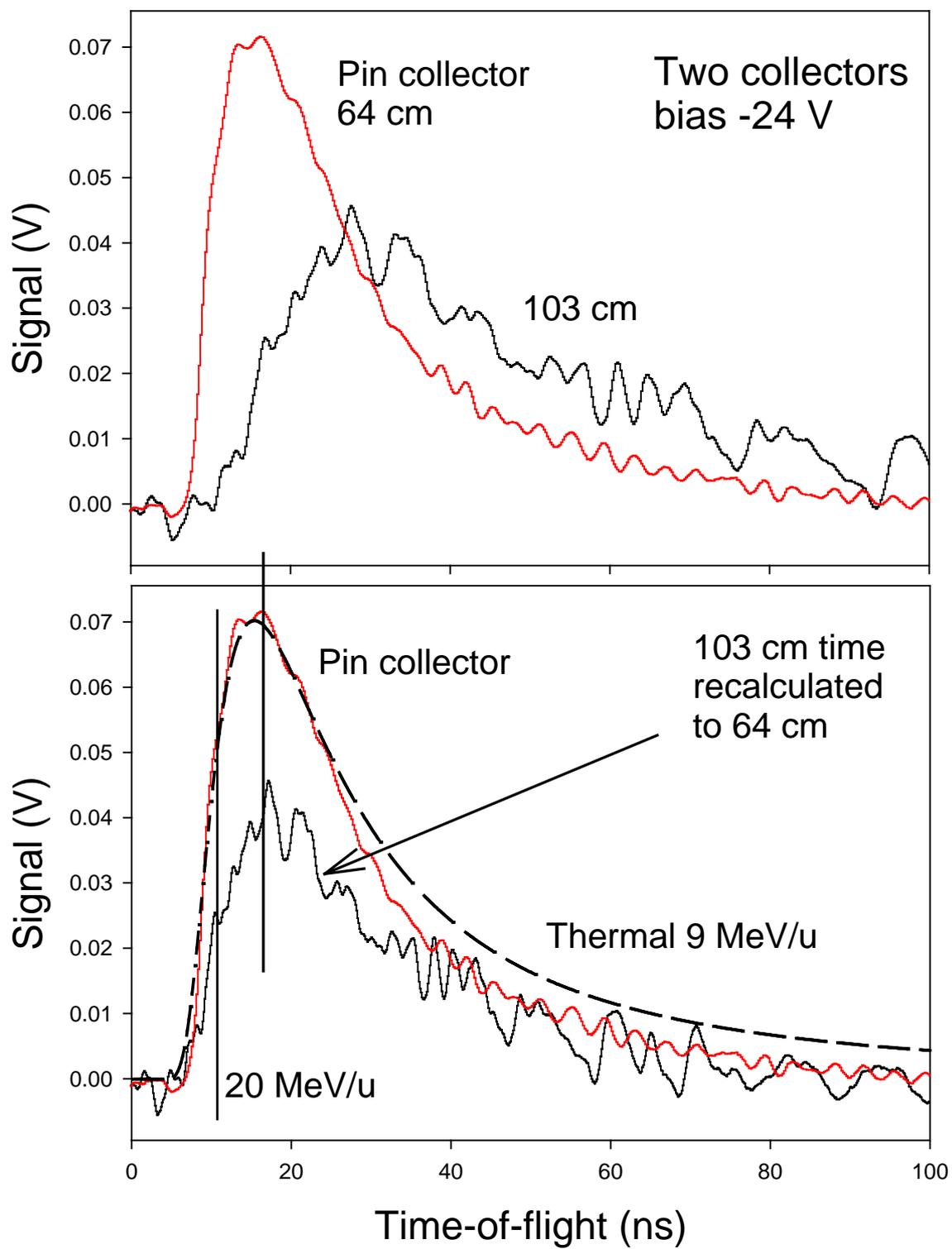

Fig. 4

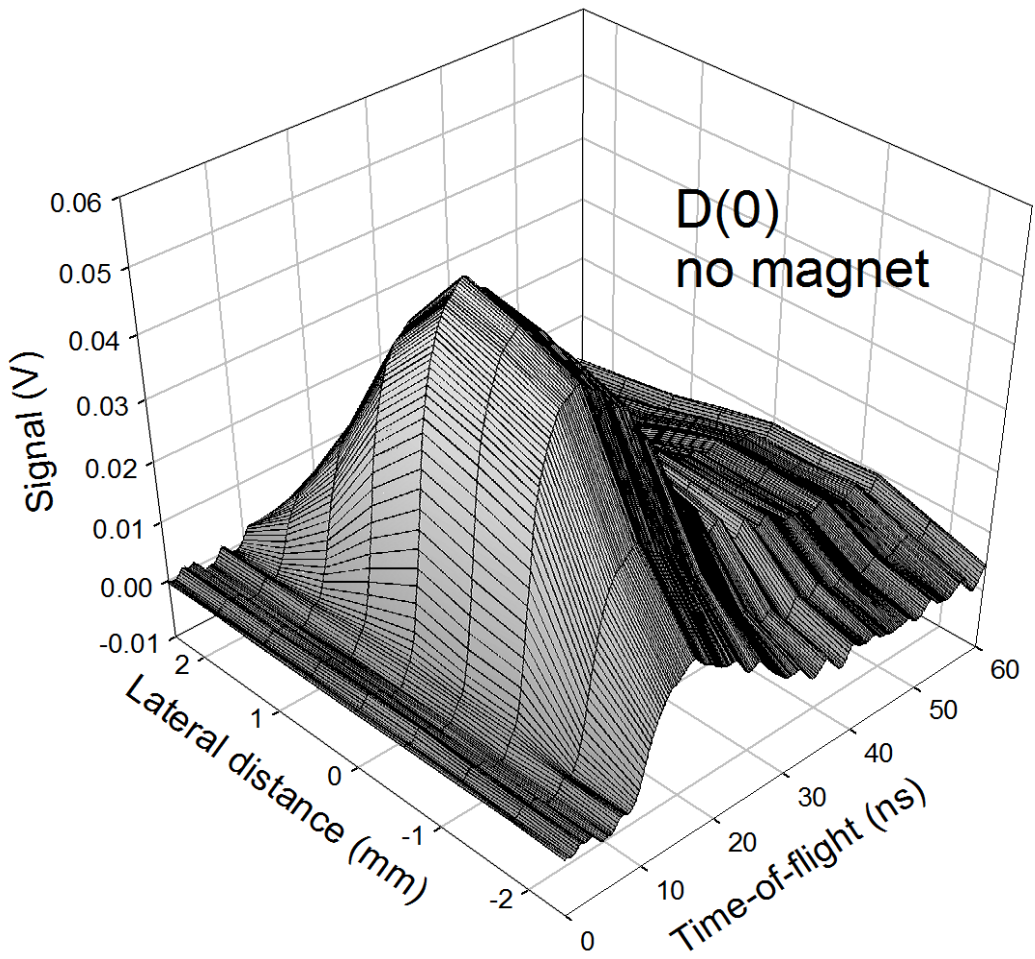

Fig. 5

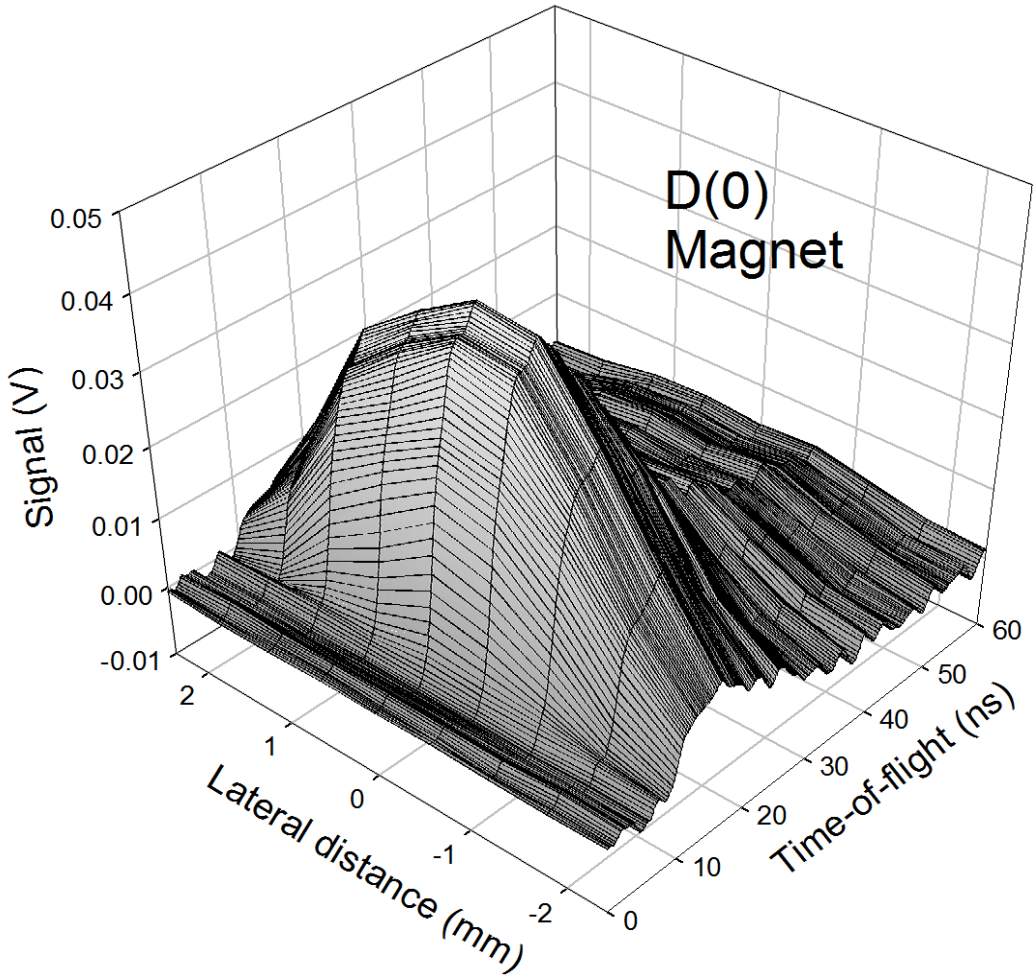

Fig. 6

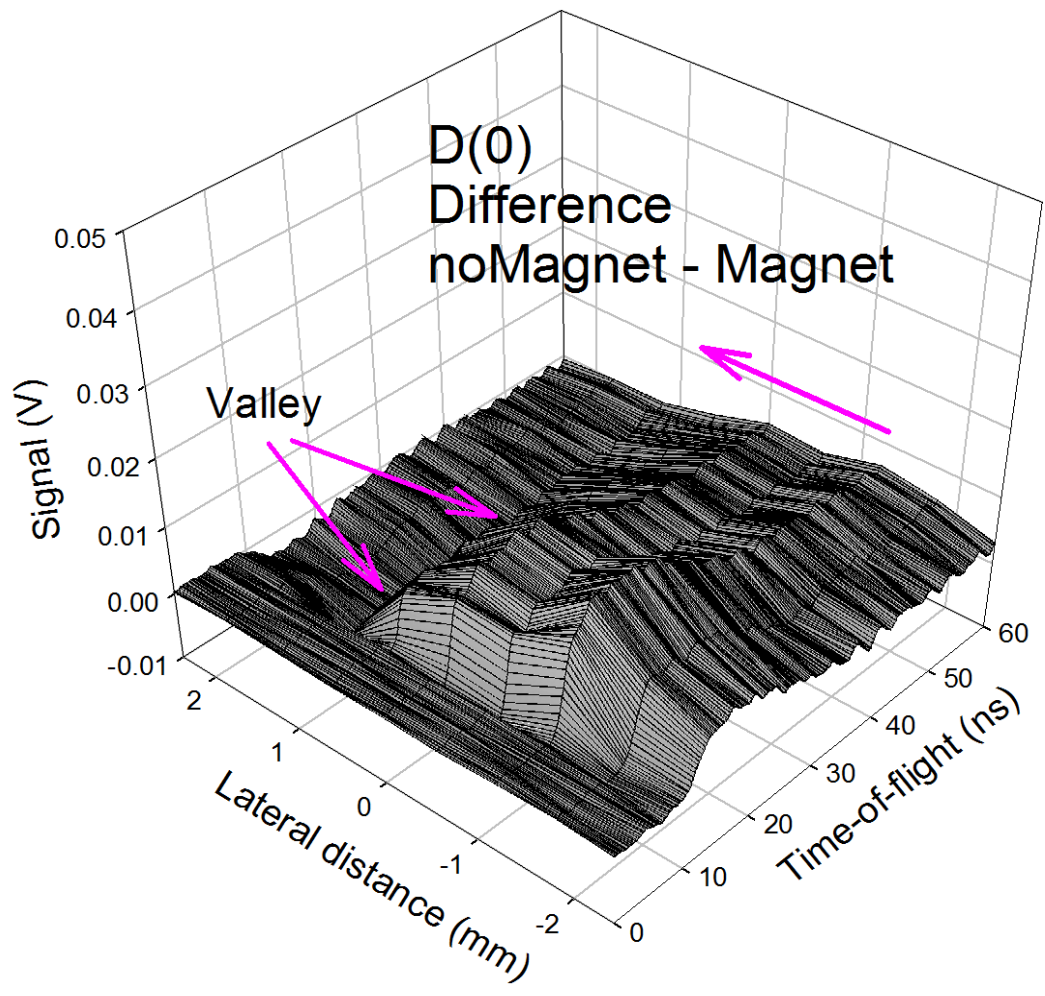

Fig. 7